# Spin coherence times of point defects in two-dimensional materials from first principles

A. Sajid[1,*] and Kristian S. Thygesen[1]

[1]CAMD and Center for Nanostructured Graphene (CNG), Department of Physics, Technical University of Denmark, 2800 Kgs. Lyngby, Denmark.

**\*Corresponding Author:** sajal@dtu.dk

**Abstract:** The spin coherence times of 69 triplet defect centers in 45 different 2D host materials are calculated using the cluster correlation expansion (CCE) method with parameters of the spin Hamiltonian obtained from density functional theory (DFT). Several of the triplets are found to exhibit extraordinarily large spin coherence times making them interesting for quantum information processing. The dependence of the spin coherence time on various factors, including the hyperfine coupling strength, the dipole-dipole coupling, and the nuclear g-factors, are systematically investigated. The analysis shows that the spin coherence time is insensitive to the atomistic details of the defect center and rather is dictated by the nuclear spin properties of the host material. Symbolic regression is then used to derive a simple expression for spin coherence time, which is validated on a test set of 55 doublet defects unseen by the regression model. The simple expression permits order-of-magnitude estimates of the spin coherence time without expensive first principles calculations.

**Introduction:**

Point defects in wide band gap semiconductors represent a promising platform for the implementation of various quantum technologies such as quantum computing and quantum sensing [1-4]. In particular, defects in atomically thin two dimensional (2D) materials have recently attracted much attention because they are easier to create, characterize, and manipulate as compared defects buried inside a bulk crystals [5,6]. Moreover, deep level defects in 2D insulators with paramagnetic spin states could offer qubit systems with distinct advantages over more conventional solid state qubit systems like quantum dots and NV centers in diamond [7]. In particular, spin qubits based on 2D crystal point defects could be easier to scale up and have better compatibility with modern semiconductor devices [8-10]. The spin coherence time $T_2$, usually measured in Hahn-echo experiments [11], is one of the most critical properties for qubits, and long spin coherence times are required for various applications[1,12,13,14].

Although there can be many sources of decoherence, e.g. instantaneous diffusion, relaxation time $T_1$, direct flip-flop, indirect flip-flop[15] etc. however, in solid state systems, in absence of other nearby paramagnetic spins and dangling bonds, the magnetic fluctuations of the surrounding nuclear spins is the dominant source of decoherence of a defect's electron spin (often referred to as central spin). Hence, for solid state systems, $T_2$ represents an upper bound on the coherence time. Defect spins in materials with a lower concentration of nuclear spins are generally expected to have longer coherence times. For this reason, 2D materials should be ideally suited as hosts of spin qubits, as compared to conventional 3D bulk materials, because of their intrinsically lower density of nuclei. However, because of the extreme thinness of 2D materials, the $T_2$ is expected to be more sensitive to the host material's environment, e.g. the substrate [16]. Such effects are not considered in the present work and all results apply to defects in the freestanding 2D materials. Studies of spin dynamics in 2D materials have so far been limited to a few systems e.g. hexagonal boron nitride (hBN) and $MoS_2$[17-19]. Given the interest in discovering novel, useful defects for quantum applications, and the importance of $T_2$ in this regard, there is a critical need for broader and more systematic studies of spin dynamics in 2D qubit systems.

In the present work, we calculate the spin coherence time of the 69 point defects with a triplet ground state currently contained in the quantum point defect database (QPOD) [20]. We find several new candidate qubit systems with very long spin coherence times, which have not been reported previously to the best of our knowledge. The convergence with respect to model parameters such as the size of spin bath, distance between two bath spins, and the level of cluster-correlation expansion are carefully studied. Our simulations show that spin coherence times are largely independent of the chemical and structural details of the point defect but are dictated by the host material's nuclear spins; specifically, their spin angular momentum, gyromagnetic ratios (g-factors), and spin-spin distances. Based on this observation, we propose a simple descriptor for $T_2$ in terms of a few features characterizing the host nuclear spins. This descriptor is validated for a set of 55 doublet systems and used to estimate the spin coherence time for all spinful defects in QPOD. All parameters entering the spin Hamiltonian and the calculated spin coherence times are available in the QPOD[20] online database.

While we were finalizing the manuscript, a similar paper by Kanai *et. al.* appeared[21] where the authors proposed an analytic expression for $T_2$ of defects in 3D host compounds with dilute nuclear spin baths. Although, there is a considerable similarity in the conclusions of the two studies, the important differences are that we present ab initio results for the $T_2$ of a range of (newly proposed) high spin defects in 2D host crystals, and the simple expression we obtain for $T_2$ applies to atomically thin materials rather than bulk materials.

**Methods:**

The Hamiltonian describing the dynamics of a triplet defect center interacting with a bath of nuclear spins in the presence of an external magnetic field takes the form

$$H_{total} = H_S + H_B + H_{S-B} \quad (1)$$

Here the terms describing the central spin and its interaction with the bath read

$$H_S = -\gamma_e \vec{B} \cdot \vec{S} + \vec{S} \cdot D \cdot \vec{S} \quad (2)$$

$$H_{S-B} = \vec{S} \cdot \sum_i \vec{A}_i \cdot \vec{I}_i \quad (3)$$

The bath Hamiltonian is given by

$$H_B = -\vec{B} \cdot \sum_i \gamma_i \vec{I}_i + H_{n-n} \quad (4)$$

where

$$H_{n-n} = \frac{\mu_B}{4\pi} \sum_{i>j} \gamma_i^n \gamma_j^n \frac{\vec{I}_i \cdot \vec{I}_j - 3(\vec{I}_i \cdot \hat{r}_{ij})(\vec{I}_j \cdot \hat{r}_{ij})}{r_{ij}^3} \quad (5)$$

In these expressions, $\gamma_e$ and $\gamma_i^n$ are the gyromagnetic ratios (g-factors) of the electron and the $i$'th nucleus, respectively. $B$, $A$, $D$ and $I$ represent the applied magnetic field, the hyperfine coupling tensor, zero field splitting tensor, and the nuclear spin operator, respectively. The first term in $H_S$ is the Zeeman term for electrons, while the second term is the zero field splitting (ZFS) term that separates the spin sub-levels of the triplet. The first term in $H_B$ is the Zeeman term for the nuclei, while the second term is the nuclear spin dipole-dipole interaction. $H_{S-B}$ describes the hyperfine (HF) interaction between the nuclear spin bath and the defect electrons. In the present work, we ignore the $S_x$ and $S_y$ components of the electron spin operator in $H_{S-B}$ so that flipping of the electron spin cannot take place. This approximation is motivated by the large difference between the electron and nuclear gyromagnetic ratios, which implies that under large magnetic fields spin flipping of the electron spin cannot occur because the associated energy cost greatly exceeds the hyperfine interaction energy[17]. Within this approximation the total Hamiltonian commutes with the $S_z$ operator and hence the Hamiltonian can be written as

$$H_{total} = H_S + H_B + m_s \sum_i (\vec{A}_i \cdot \vec{I}_i)_z \quad (6)$$

where $m_s$ is the magnetic quantum number of the central spin. We note that the quadrupole interaction term is ignored here. Previous studies have shown that its inclusion may increase $T_2$ for some materials. Ref. [21] found an increase in $T_2$ by about 30% for $WS_2$ while Ref. [16] observed a change in $T_2$ from 2.2 to 4.1 ms (an increase of about 50%) for a particular spin defect in $MoS_2$ upon inclusion of the quadrupole term. Hence, our results represent a lower bound on the $T_2$. The decoherence of the central spin triplet is studied by considering the central spin and the nuclear spins (environmental bath) as a closed quantum system. In practice, the nuclear bath is represented by all nuclear spins located within a radius of 50 Å from the central spin. This value of bath radius, ensures converged $T_2$, as discussed later in the text. The nuclear spin ($\vec{I}_i$) and g-factor ($\gamma_i$) of a given atom are chosen according to the natural abundance of the isotopes of the atomic species. The combined qubit and bath system is initially prepared in a product state of the form,

$$|\psi(0)\rangle = \frac{1}{\sqrt{2}}(|1\rangle + |0\rangle) \otimes |B(0)\rangle \quad (7)$$

where, $|1\rangle$ and $|0\rangle$ represent the $m_s = +1$ and $m_s = -1$ states of the triplet, respectively, and $|B(0)\rangle$ is the state of the spin bath at $t = 0$. At any later time $t$ the bath state entangles with the qubit states and the combined state is given by

$$|\psi(t)\rangle = \frac{1}{\sqrt{2}}(|1\rangle \otimes |B^{(1)}(t)\rangle + |0\rangle \otimes |B^{(0)}(t)\rangle) \qquad (8)$$

where $|B^{(0)}(t)\rangle$ and $|B^{(1)}(t)\rangle$ are the bath states at time $t$ conditioned on the state of the qubit. The phase information of the central spin at an arbitrary time $t$ is encoded in the off-diagonal elements of the reduced density matrix, $\vec{\rho_s}$, which in turn equals the overlap of the two bath states. The coherence function $\mathcal{L}(t)$ describes the loss of the relative phase of $|0\rangle$ and $|1\rangle$, and is defined by

$$\mathcal{L}(t) = \frac{\langle 1|\vec{\rho_s}(t)|0\rangle}{\langle 1|\vec{\rho_s}(0)|0\rangle} = 2\langle B^{(1)}(t)|B^{(0)}(t)\rangle \qquad (9)$$

For a bath size of a few hundred spins or more, the coherence function can be efficiently calculated within a Cluster Correlation Expansion (CCE) scheme [22-24]. The key idea of the CCE method is that the nuclear spinbath induced decoherence of electron spin can be factorized into set of irreducible contributions from spin bath clusters, e.g. (for clusters with up to two spins)

$$\mathcal{L}(t) = \prod_i \mathcal{L}_i(t) \prod_{ij} \mathcal{L}_{ij}(t) \qquad (10)$$

where $\mathcal{L}_i(t)$ is the contribution of the single bath spin $i$ and $\mathcal{L}_{ij}(t)$ is the irreducible contribution of the spin pairs $ij$. The maximum size of the cluster included in the expansion determines the order of the CCE approximation. At the CCE-1 level each nuclear spin is treated independently and it interacts with the electron spin through the hyperfine coupling. At CCE-2 and CCE-3 levels, there are two and three spins within a correlation cluster, respectively. For any nuclear spin bath, CCE expansion provides the exact solution when the expansion includes the largest possible nuclear spin clusters (i.e. the entire nuclear spin bath) i.e. for a bath containing three distinct nuclear spins, CCE-3 provides an exact solution for the electron spin coherence. In conventional CCE approach, the total Hamiltonian of the system is conditioned onto the qubit levels, while in generalized CCE (gCCE) approach, central spin degrees of freedom are directly included into each nuclear spin cluster.

In the present work, we use the PyCCE[25] implementation of the conventional CCE method to calculate $\mathcal{L}(t)$. We limited ourselves to CCE order 2 (CCE-2) as our convergence studies show there is essentially no change in $\mathcal{L}(t)$ when going from CCE-2 to CCE-3.

The atomic structures of the defect systems, the HF couplings and the ZFS tensors were obtained from the QPOD database (we note that in the present work, the ZFS does not affect the calculated $\mathcal{L}(t)$)[20]. All the data in the QPOD database was generated by DFT calculations performed using GPAW electronic structure code[26], which is based on the projector-augmented wave ( PAW) method. All DFT calculations employed a plane wave basis set with 800 eV cut off, a k-point density of 6 Å (12 Å) for structural relaxations (property evaluations), and the PBE xc-functional [27]. The supercell was kept fixed during relaxation and atoms were fully relaxed until forces were below 0.01 eVÅ$^{-1}$. The size of the supercell was chosen to ensure that all point defects were separated by at least 15 Å.

Within CCE approach the size of the spin bath was limited to a sphere of radius $R_{\text{bath}} = 50$Å from the central spin while the maximum distance between two nuclear spins forming an irreducible pair was kept at $r_{\text{dipole}} = 15$Å. The convergence of $\mathcal{L}(t)$ w.r.t. these parameters is discussed later. A magnetic field of 5 Tesla oriented along the z-direction was applied in all the calculations, although it has been recently shown that the hetero-nuclear spin baths are decoupled in most of the compounds under standard experimentally easy achievable magnetic field as low as a few mili Tesla[21]. We note that there could be some dependence of $T_2$ on the direction of the magnetic field for defects with strongly anisotropic zero field splitting tensors [28]. In order to capture such effects, one would have to include the zero field splitting in the spin Hamiltonian and employ the gCCE approach for the computation of $T_2$. This is, however, beyond the scope of current work. We averaged the calculated $\mathcal{L}(t)$ over 100 different randomly generated spatial realizations of nuclear spins and all the results presented herein represent such an ensemble average.

The symbolic regression[29,30] was performed using the FEYN[31] package. The space of mathematical functions used to build the model ranges from addition, multiplication and squaring, to more complex functions such as the natural logarithm. The mean square error was used as loss function and a penalty term was added to regulate the number of features in the model according to the Bayesian Information Criterion (BIC)[32].

**Results and discussions:**

We first explore the convergence of the coherence function $\mathcal{L}(t)$ with respect to the important model parameters, namely the bath size $R_{\text{bath}}$, the maximum distance between two bath spins $r_{\text{dipole}}$, and the order of the cluster-correlation expansion. To that end, we systematically varied these parameters and studied the dependence of the coherence function for the three representative triplet defects WAu$_2$O$_4$-W$_{\text{Au}}^0$, Pd$_2$S$_4$-v$_{\text{Pd}}^0$, and MoAu$_2$O$_4$-Mo$_{\text{Au}}^{+1}$ (throughout this manuscript we use the naming convention X-Y where X is the chemical formula of the host material and Y denotes the defect. We use the notation v$_{\text{A}}^q$ for an A-vacancy defect in charge state q).

In Fig. 1 We show the convergence of the coherence function $\mathcal{L}(t)$ with respect to $R_{\text{bath}}$, $r_{\text{dipole}}$ and the order of the CCE for each of the three test systems. It is evident from Fig. 1 that the coherence function is well converged for $R_{\text{bath}} = 50$Å, $r_{\text{dipole}} = 15$ Å, and a CCE at order 2. Consequently, these parameter values were used for all the simulations.

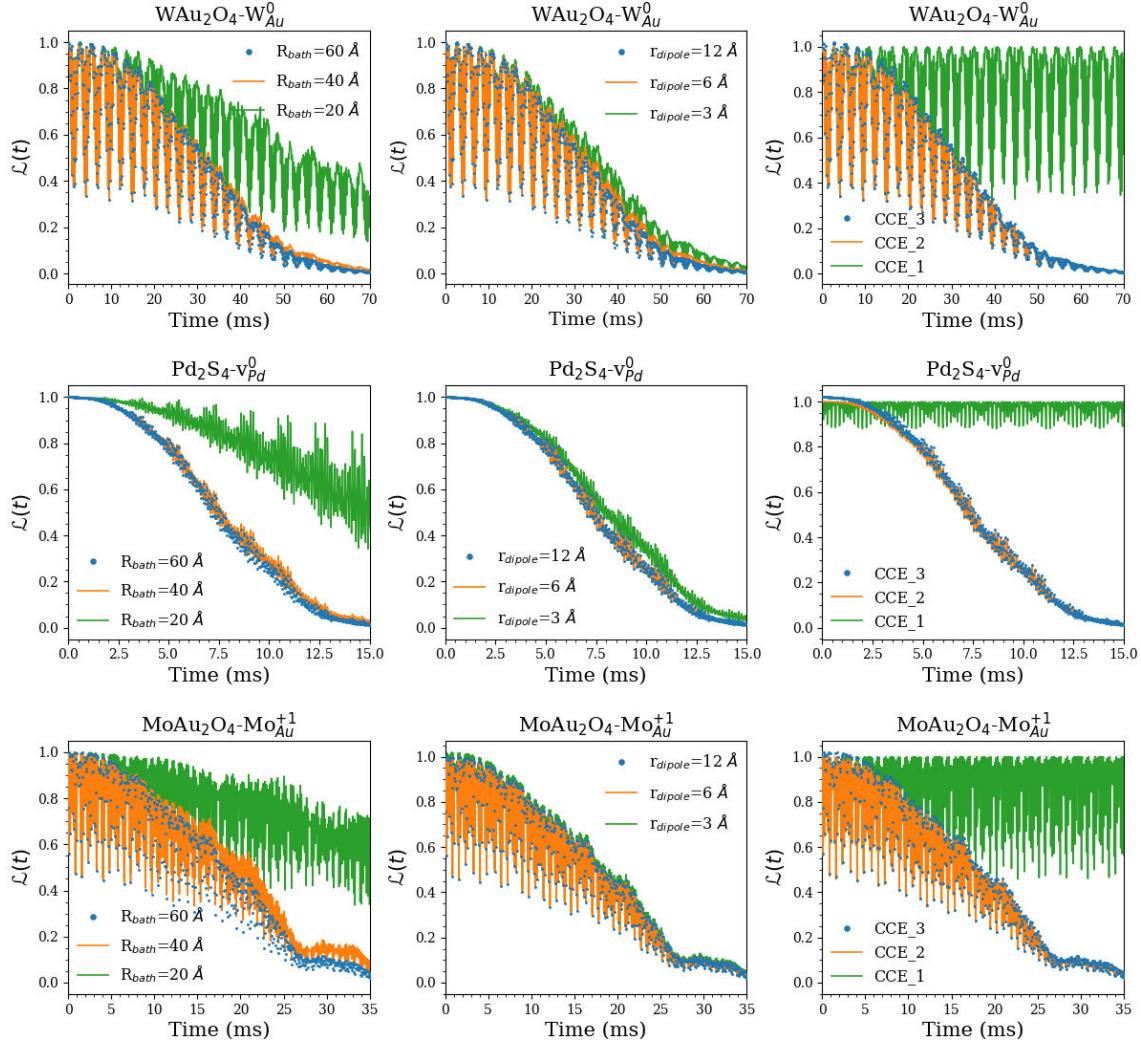

**Fig 1.** Calculated coherence function, $\mathcal{L}(t)$ vs time for different values of the parameters $R_{\text{bath}}$, $r_{\text{dipole}}$, and order of the CCE approximation for the $WAu_2O_4\text{-}W_{Au}^{0}$, $Pd_2S_4\text{-}v_{Pd}^{0}$ and $MoAu_2O_4\text{-}Mo_{Au}^{+1}$ defects (top, middle and bottom panel, respectively).

We note that the fast modulations in $\mathcal{L}(t)$ are due to non-zero probability of the nuclear spin resonance (or $m_I = \pm 1$) transitions called electron spin echo envelop modulations (ESEEM)[33,34]. One can note that ESEEMs are captured already by calculations at order CCE-1 at short time scales, in agreement with a previous study[25]. The overall decay of $\mathcal{L}(t)$ is well converged at level CCE-2, and we therefore limit ourselves to CCE-2 in the present work. This choice saves substantial computer time and ensures a reasonable accuracy for the calculation of coherence function.

Using the parameters determined from the convergence test, we calculated the coherence function of 69 triplet defects from the QPOD database. From the coherence functions, the spin coherence time, or Hahn echo time $T_2$, was obtained by fitting the coherence function to the stretched exponential, $\exp(-t/T_2)^n$, with coherence time $T_2$ and stretching exponent $n$. An example is shown in Fig. 2 for the example of a sulphur vacancy in $MoS_2$ in charge state -2 (i.e. $MoS_2$-$v_S^{-2}$).

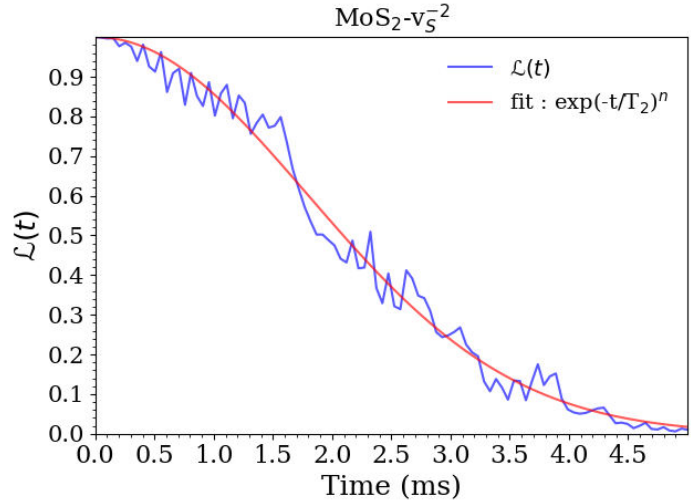

**Fig 2.** The calculated coherence function for a sulphur vacancy in $MoS_2$ in charge state -2. The spin coherence time is extracted by fitting the coherence function by the stretched exponential function $\exp(-t/T_2)^n$. The extracted values of $T_2$ and $n$ are 2.51 ms and 2.03, respectively.

The calculated spin coherence times $T_2$ for all the triplet defect systems studied in the present work are shown in Fig 3. The colored sections indicate defects in the same host material. One can immediately conclude that $T_2$ is mainly a property of the host material as different defects embedded within the same host exhibit very similar coherence times. This is because $T_2$ is determined by the magnetic fluctuations of thousands of nuclear spins and hence is independent of the details of the defect structure, as long as the spin density of the defect is well localized. The results also show that the hyperfine coupling strength (which is a property of the central spin that varies significantly for different defects in the same host material) plays a minor role for the spin dynamics, in general. The calculated $T_2$ for all the defect species are listed in the QPOD database[20].

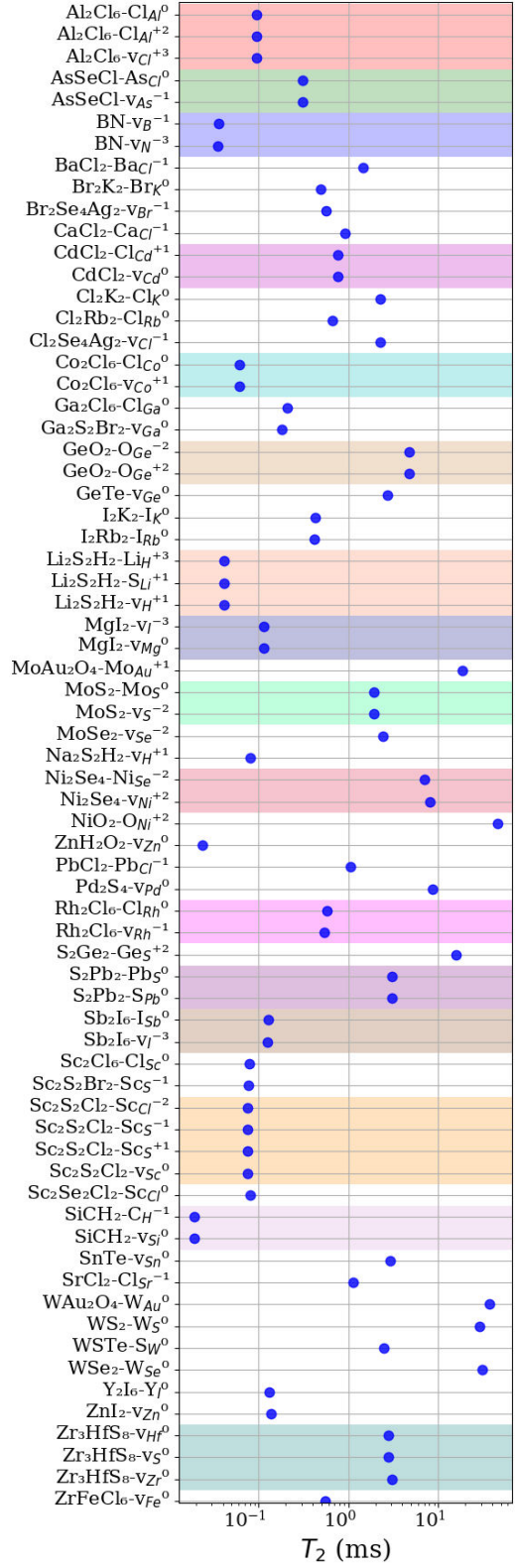

**Fig 3.** The calculated spin-coherence times $T_2$ for 69 triplet defect systems within 48 distinct 2D hosts. The x-axis lists the triplet defect systems, while $T_2$ are plotted on a log-scale. Different defects within the same hosts are collected and highlighted by different colors.

Another thing to note is the exceptionally long spin coherence times of 29ms (30ms), 37ms, 46ms, and 18ms obtained for the defects in the 2D materials $WS_2$ ($WSe_2$), $WAu_2O_4$, $NiO_2$, and $MoAu_2O_4$, respectively. The high values of the spin coherence time within these compounds can be attributed to the low concentration of spinful isotopes of the involved elements and/or their low g-factors (see table S1). The coherence times can be further increased through the process of isotopic purification [18,35]. Point defect species embedded within host materials with such long spin coherence times are very attractive for qubit applications. Furthermore, these hosts have relatively high bandgaps (HSE06) i.e. 2.06 eV (and 1.73 eV), 4.44 eV, 3.31 eV and 4.11 eV, for $WS_2$ (and $WSe_2$), $WAu_2O_4$, $NiO_2$, and $MoAu_2O_4$, respectively[36]. The other relevant thermodynamic and magneto-optical properties e.g. defect formation energies, charge transition levels, Fermi level positions, equilibrium defect and carrier concentrations, transition dipole moments, hyperfine coupling, and zero-field splitting of the particular triplet, vacancy and/or substitutional defects, studied in the present work for these hosts are presented in the QPOD database[20].

We also note that our calculated spin coherence times for known defect species in hexagonal boron-nitride (hBN) and $MoS_2$ (in particular the boron vacancy in the -1 charge state in hBN and the sulphur vacancy in the -2 charge state in $MoS_2$) are in very good agreement with previous studies [17]. Moreover, the calculated coherence time of 29 ms for $WS_2$ matches reasonably well with another recent theoretical study[16]. It is interesting to note that the $T_2$ for bulk $WS_2$ is 13.6ms [16] and is three times lower than the value for 2D $WS_2$, consistent with the higher nuclear spin density in 3D. This effect of dimensionality on $T_2$ (2D vs 3D) is consistent with the previous study for other materials[17].

We now move on to discuss the dominant factors that govern $T_2$ for a defect in a given host material. Crystal geometry, in particular the interatomic distances, is expected to play a significant role as the decoherence of the central spin is caused by flip-flop transitions of the nuclear spin bath, and the dipole-dipole interaction driving these transitions depends on distance as $1/r_{ij}^3$. Another factor influencing $T_2$ is the nuclear spin of the atoms of the host material, as the dipole-dipole interaction is directly proportional to the product of nuclear spins, $I_iI_j$. The concentration, or natural abundance, of the nuclear isotopes with non-zero spin should be another important factor. Indeed, host systems like BN, $SiCH_2$, and $ZnH_2O_2$, which all have large concentrations of nuclear spins (e.g. natural abundance of the $B^{11}$ isotope with spin $I=3/2$ is 80% while it is 99.99% for the $H^1$ isotope with $I=1/2$), have very short spin coherence times, as can be seen in Fig. 3. Finally, as the dipolar interaction between nuclear spins $i$ and $j$ is proportional to $\gamma_i^n \gamma_j^n$, the nuclear g-factors are also expected to be important. Another point to note is the similarity between the $T_2$ values of $WS_2$ and $WSe_2$ (and $MoS_2$ and $MoSe_2$). Although Se has a higher concentration of nuclear spin isotopes and a larger g-factor than S (see table S1), this is counter balanced by the smaller magnitude of the nuclear spin of Se compared to S (0.5$\hbar$ versus 1.5$\hbar$).

While all of these factors are expected to be important for $T_2$, it will not be possible in practice to optimize them all separately. Consequently, for the purpose of selecting or designing good host material for qubit applications, it is important to establish how the different parameters play together to determine $T_2$. To this end, we show in Fig. 4 the correlation between the calculated $T_2$ and each of the four key parameters: average of the non-zero nuclear spins ($\langle I \rangle$), average concentration of non-zero nuclear spins ($\langle \rho \rangle$), average g-factor of nuclei with non-zero spin ($\langle \gamma^n \rangle$) and average nearest neighbor distance between nuclear spins ($\langle d_{nn} \rangle$). It is clear that none of the parameters alone can explain the trend in $T_2$. The form of the dipole-dipole interaction in Eq. (5) suggests the following simple descriptor for the spin coherence time

$$D^{ct} = \frac{1}{N} \frac{1}{\bar{d}_{nn}} \sum_{i=1}^{N} \gamma_i^n \rho_i I_i \qquad (11)$$

where the summation $i$ run over the spinful isotopes of the elements of the host material, $\bar{d}_{nn}$ is the average nearest neighbor distance of nuclear spins, and $\rho_i$ is the natural concentration of the spinful isotopes.

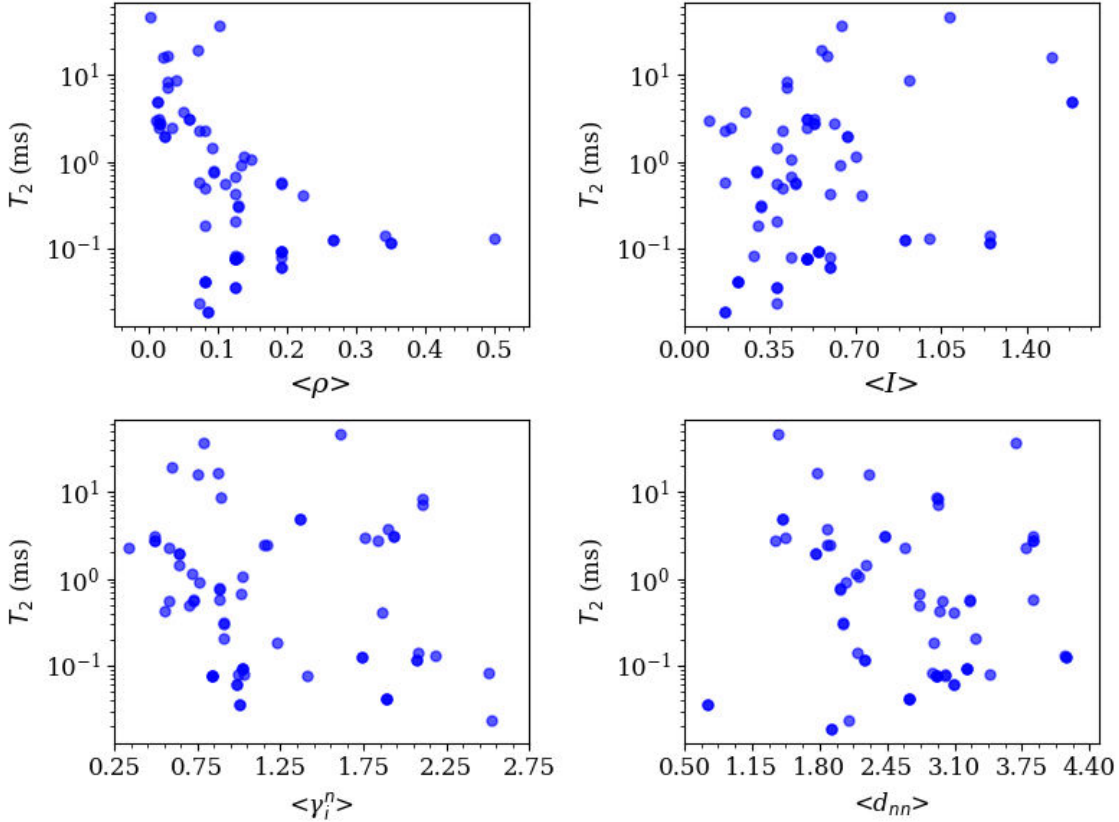

**Fig 4**. Spin coherence time, $T_2$, versus average natural abundance/concentration of spinful isotopes ($\langle \rho \rangle = \frac{1}{N}\sum_{i=1}^{N} \rho_i$), average nuclear spin ($\langle I \rangle = \frac{1}{N}\sum_{i=1}^{N} I_i$), average nuclear g-factors ($\langle \gamma^n \rangle = \frac{1}{N}\sum_{i=1}^{N} \gamma_i^n$) and average nearest neighbor nuclear spin distance in the host material ($\langle d_{nn} \rangle$). It can be observed that there is no clear correlation between $T_2$ and any of the four descriptors individually.

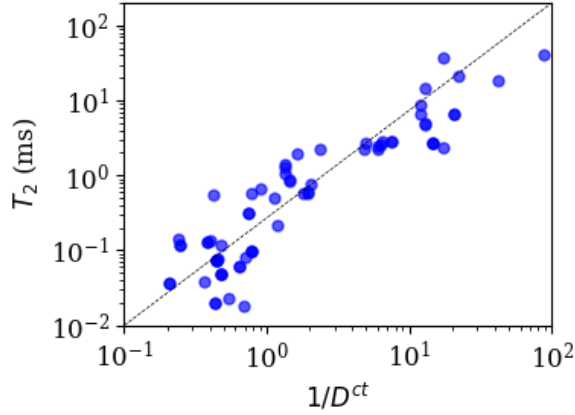

**Fig 5.** The calculated $T_2$ versus the descriptor $1/D^{ct}$ in the units of (radian-T$^{-1}$s$^{-1}$m$^{-3}$)$^{-1}$ (Eq. 11). A clear correlation can be seen between $T_2$ and $D^{ct}$.

With the aim of obtaining an even better $T_2$ descriptor, we applied symbolic regression on the data (training) set of 69 triplets. Our primitive feature set consists of the host material features: $\langle I \rangle$, $\langle \rho \rangle$, $\langle \gamma^n \rangle$, $\langle d_{nn} \rangle$ and $D^{ct}$. We generate a total of 11159 features by performing various mathematical operations on the set of primitive features. We then perform regression of $\log T_2$ in the large feature space using the Bayesian Information Criterion (BIC)[32] to penalize models with many features. The best fit model depending on up to two features is given in Eq. 12. Figure 6 shows the true versus predicted $T_2$ for both the training set (69 triplets) and test set (55 doublets). One can note from Eq. 12 that a significant weight is assigned by the regression method to the descriptor $D^{ct}$. The detailed training metrics are listed in the supplementary information[37] Table S3. The mean relative error is 35.7% for the training set and 41.3% for the test set. These errors are in fact quite low considering that the $T_2$ values of the data set varies over three orders of magnitude.

The fitting model details, plots and metrics for the spin coherence times, for a data set which does not include $D^{ct}$ as a feature is shown in the supplementary information[37] Eq. S1, Fig S4 and table S3, respectively. One can clearly see that the machine learning predictions of spin coherence time only gets worse by not including $D^{ct}$ as a feature. We further tried to fill up our feature space with many different combinations of the four features described above and found that symbolic regression do not include those complex features during the training process.

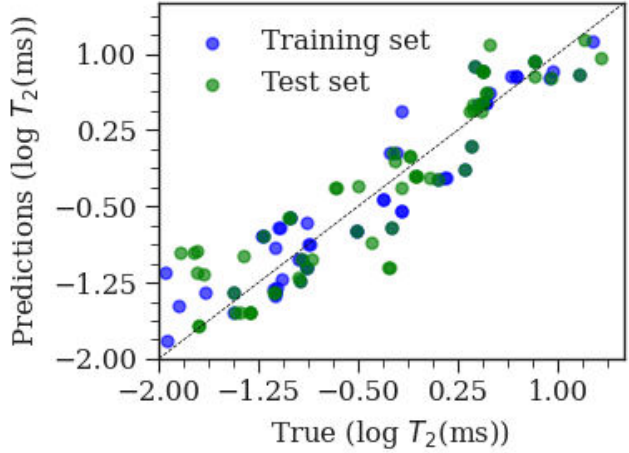

**Fig 6.** The true versus predicted coherence times (log $T_2$ (ms)) for the training set (blue) and test set (green). The line of equality is drawn to give an idea of accuracy of prediction.

$$\log T_2 = -0.92 \log\langle d_{nn}\rangle + 0.78\log\langle\rho\rangle - 1.67 \log D^{ct} + 0.52 \qquad (12)$$

**Conclusion:**

We have calculated the spin coherence times ($T_2$) of a large set of point defects in different 2D host materials and found new systems with exceptionally large spin coherence times. In particular, defects in WS$_2$ (and WSe$_2$), WAu$_2$O$_4$, NiO$_2$, and MoAu$_2$O$_4$ all have $T_2$ above 15ms. Based on our results we conclude that the spin coherence times is a property of the host materials and is insensitive to the atomistic details of the defect center. We have performed detailed investigations of how various elementary host specific properties influence the spin coherence time of defect spin centers. On that basis we propose a simple descriptor that correlates very well with the coherence time and can be used to identify crystals that could host defects with long spin coherence time without resorting to expensive first principles calculations. Our work provides insight into the coherence of defect spin qubits in 2D materials.

**Author contributions:**

S. A. conceived the idea, performed the calculations and analysis. K. S. T. supervised the project. All authors discussed and modified the manuscript.

**Acknowledgment:**

This work is supported by Novo Nordisk Foundation Challenge Programme 2021: Smart nanomaterials for applications in life-science, BIOMAG Grant NNF21OC0066526. We also acknowledge funding from the European Research Council (ERC) under the European Union's Horizon 2020 research and innovation program Grant No. 773122 (LIMA). K. S. T. is a Villum Investigator supported by VILLUM FONDEN (grant no. 37789).

**Competing Interests:** The authors declare no competing interests.